\documentclass[11pt]{article}
\usepackage{amsmath,amssymb}
\usepackage{framed}

\begin{document}

\setcounter{page}{1}

\pagestyle{plain}

\begin{center}
\Large{\bf Visible Energy Alternative to Dark Energy}\\
\small \vspace{1cm} {Maryam
Roushan}$^{}$\footnote{m.roushan@umz.ac.ir},\quad {Narges
Rashidi}$^{}$\footnote{n.rashidi@umz.ac.ir }\quad and \quad {Kourosh
Nozari}$^{}$\footnote{knozari@umz.ac.ir (Corresponding
Author)} \\
\vspace{0.5cm} $^{}$Department of Theoretical Physics, Faculty of
Science, University of Mazandaran,\\ P. O. Box 47416-95447, Babolsar, Iran\\
ICRANet-Mazandaran, University of Mazandaran, P. O. Box 47416-95447,
Babolsar, Iran\\
\end{center}

\begin{abstract}
Quantum gravitational effects usually are assumed to be important on
small scale (Planck scale), but actually these effects are also very
significant on large (cosmological) scales. It is recognized that in
curved spacetime, the existence of a minimal measurable momentum is
inevitable. In this paper, we study thermodynamic properties of the
late time universe in the presence of a minimal measurable momentum
cutoff that encodes infra-red modification of the underlying field
theory. In this regard, we consider a non-relativistic regime and
show that the existence of a minimal measurable momentum in the very
essence of the theory leads to accelerating expansion of the
universe, which can be interpreted as an alternative to Dark Energy.
The
universe in this model has experienced the phantom line crossing in the near past.
\end{abstract}

\section{Introduction}

It is well known that alternative scenarios to quantum gravity
proposal  demonstrate the existence of an inherent ultraviolet (UV)
cutoff in
nature~\cite{Veneziano1986,Amati1989,Konishi1990,Kempf1995,Roushan2016,Roushan2019,Camelia2002}.
The effects of this natural cutoff are significant at short
distances (Planck scales). On the other hand, the gravitational
effects are not only noticeable at short distances; they are also
remarkable at large distances (cosmological scales). The  curvature
of the spacetime at large scales refers to an infrared (IR)
cutoff~\cite{Kempf1996}. It should be emphasized that gravity is a
nonlinear field theory and the particle moving along a curved
spacetime geodesic affects the geodesic itself. At the same time,
the residual effects of gravity don't allow the particle's energy
and therefore its momentum to be zero. Therefore, it is logically
consistent to assume that there is a minimal uncertainty in
momentum.

In this situation along with minimal momentum effects (which are
inherent features of spacetime), many of the basic concepts of
standard quantum mechanics lose their degree of credibility. There
is even a question in this background: Is essentially spacetime
discrete or continuous? This point has been widely studied in recent
years~\cite{Kempf1995,Kempf1996,Roushan2014,
Roushan2018}. Indeed,
the existence of an uncertainty gap in the configuration of
spacetime causes the fabric of spacetime to be discrete. Therefore,
position and momentum operators are also affected by this feature of
spacetime so that to preserve the symmetry of the theory, the
corresponding operators must be redefined. For this purpose, we use
new position and momentum operators~\cite{Roushan2020}.

Another property of this background is that the plane wave is
incomprehensible in curved spaces. This implies that there is a
bound on the resolution of momenta which can be
characterized~\cite{Kempf1996,Kempf1994}. We will continue with the
prediction that this can be represented as a nonzero minimal
uncertainty in the momentum measurement, where we imply the ordinary
description of the uncertainty in an observable. The mechanism of
the existence of a minimal momentum causes the geometric corrections
of spacetime structure. More precisely, it leads to noncommutative
spacetime. Also, minimal uncertainty in the momentum can be
considered as the regularization of the infra-red (IR) effect in the
quantum field theory. In recent years, some attempts are dedicated
to probing the role of quantum gravity at the cosmological scales
and especially on the late time cosmological
dynamics~\cite{Elizalde1994,Page2010}. In this regard, we have
investigated the thermodynamics of the universe in our previous
work~\cite{Roushan2020} for particles with modified relativistic
energy in the presence of natural cutoffs.

Observational data confirms that our universe is currently
undergoing an accelerating phase of expansion. However, Einstein's
field equations with the standard matters can not describe this
accelerating expansion. In this regard, the cosmologists have
suggested that Einstein's field equations should be modified. This
modification can be performed in two major road maps. One way is to
modify the energy-momentum side of the field equations by
considering some sort of matter fields, such as scalar fields (as
dark energy components)~\cite{Copeland2006}. Another way is to
modify the geometric side of the field equations, like the $f(R)$
models (as dark geometry components)~\cite{Nojiri2020}. There are
also some other alternatives to explain late time cosmic speed up
and large scale structure formation. For instance, the authors in
Ref.~\cite{Capo2017Noether} have studied cosmology in a teleparallel
modified gravity framework. In this regard, they have considered a
different spacetime without curvature among a nonzero torsion in
contrast to General Relativity. This model has been investigated by
the Noether symmetry method and belongs to the second category.
Also, in Ref.~\cite{CapoMir2017}, the authors have studied nonlocal
deformations of teleparallel gravity including its cosmological
solutions. They have explained that the nonlocal deformations of
teleparallel gravity are inspired by quantum gravitational effects.
In addition, in Ref.~\cite{Capo2020Addressing}, the authors have
considered the effects of the uncertainty principle on the
reliability of cosmological measurements. In this regard, they have
used the relation between the Compton wavelength and the rest mass
of a particle. The authors have supposed the reduced wavelength as a
length measurement, and the definition of the rest mass as
$m_{0}=\frac{\Delta p}{c}$. In this way, by using the luminosity
distance, they have found an equation for mass in terms of the
reduced Planck constant, redshift, deceleration, and Hubble
parameter. It has been shown that the uncertainty on the photon
mass, can be the reason for the $H_{0}$ tension. In another work,
the authors of Ref.~\cite{Acunzo2021Capo}, have considered the
models based on the Integral Kernel Theories of Gravity, which is a
non-local straightforward extension of $f(R)$ gravity. In these
models, it is possible to interpret dark energy as a geometric
contribution. The authors of this paper have studied the non-local
gravity models by considering the so-called Noether Symmetry
Approach and have found some interesting analytic cosmological
solutions. Another aspect of non-local gravity has been studied in
Ref.~\cite{Capo2021Logarithmic}, where the authors have studied the
large-scale structure. It has been shown that the logarithmic
correction modifies the gravitational partition function and impacts
on properties of the gravitational clustering.

Now, we want to describe the accelerating expansion of the universe
by using a new distinct method without any dark component (neither
dark energy nor dark geometry). We consider this fact that our
universe contains two broad standard classes of particles. One class
is relativistic particles, which is relevant to the early universe
and the second class is non-relativistic particles, which concerns
the late time universe (the thermodynamics of the universe infers
that non-relativistic energies dominate the late time universe).
Here, we focus on the non-relativistic class of the particles in the
presence of a minimal measurable momentum at the late time universe.
In this regard, the study of the thermodynamics of the universe can
enhance our understanding of the accelerating phase of the universe
in this regime. In this paper, we are seeking to describe the
expansion of the universe using thermodynamic approaches. Our
motivation for proposing this scenario is that various theories of
cosmology have been expanded mainly from a thermodynamic
perspective. In this connection, we will answer the question that
whether considering the infrared cutoff may be a new approach to
explain the accelerating expansion of the universe. The important
point in this work is the fact that we consider the quantum
gravitational corrections at the low energy limit of the quantum
gravity, corresponding to the late time universe. We show that
considering this quantum correction, even in its weak field limit,
is enough to explain the late time acceleration fascinatingly. In
this perspective, there is no need for unknown, dark, and obscure
sources to explain late time speed up. This is because that we are
dealing with the energy of non-relativistic particles (standard
model particles) which are known and transparent. The energy does
not have the ambiguities of dark energy, but can be a simple and
visible alternative for it. For this reason, in this article, we
will use the phrase ``Visible Energy" against ``Dark Energy". Note
that, this perspective is studied for the first time in this context
and we think it opens new windows and sheds light on a better
understanding of how the universe works at the late time.

\section{Gravitational effects at the large distances}
Generically, there is no concept of a plane wave in the curved
spaces at large distances. This consequence can be characterized as
a property of spacetime. In other words, it can refer to
noncommutative geometry. Also, it should be noticed that the
fundamental properties of plane waves are eigenfunctions of the
momentum operator and therefore the momentum operator definition
should be modified in the noncommutative phase space. In fact, there
exists a finite minimal uncertainty in momenta in curved
space~\cite{Hinrichsen1996,Mirza2009}. In Ref.~\cite{Kempf1996}, the
author demonstrated IR regularization regarding noncommutative
geometries, which suggests the existence of minimal uncertainties in
the momenta. From a technical perspective, the representation theory
loses its credibility due to infrared regularization, such that we
have to use distinct Hilbert space representations. The
interrelations between the operators of noncommutative phase space
have first been analyzed in Ref.~\cite{A.Kempf1994}. Note that, by
considering the correction, the momenta lose their credibility as
generators of translation on the flat space. In specific
circumstances, the momenta can generate translation on the curved
space. In fact, there is a direct relationship between the existence
of minimal uncertainty in the momenta and the nonexistence of the
flat (plan) wave~\cite{Kempf1996}.

In this paper, we address noncommutative geometric corrections
through the canonical commutation relation as
follows~\cite{Roushan2020}
\begin{equation}\label{eq1}
[x_i,p_j]=i\hbar(\delta_{ij}+\eta_{ijkl}x^kx^l+...)\,.
\end{equation}
In this noncommutative phase space, the position operators are
allowed to commute with each other $[x_i,x_j ]=0 $, and only the
momenta are noncommutative $ [p_i,p_j]\neq0$. In this situation, the
emergence of a nonzero minimal uncertainty in the momenta can
regularize infrared divergencies. Another important issue about this
discrete spacetime is that the position and momentum operators will
find new definitions. Our suggestion for redefining these operators
is as follows
\begin{eqnarray}\label{eq2}
X_i=x_{i}\,,\quad P_i=p_i(1+\eta x^2)\,,
\end{eqnarray}
where $\eta$ is a small parameter that corresponds to the infrared
effect of this hypothesis. Also, $x$ and $p$ are the position and
momentum operators of the Heisenberg algebra of the standard quantum
mechanics.

\section{Quantum gravitational effects on the Maxwell-Boltzmann statistics of the late time universe}

Observations confirm that our universe is in an accelerating
expansion phase today. Since this accelerating expansion was
discovered, several investigations have been done in this area, such
as various types of models with scalar fields (dark energy
models)~\cite{Copeland2006} and $f(R)$ gravity models (dark geometry
models)~\cite{Nojiri2020}. In this paper, we seek to investigate
this phase of accelerating expansion by considering the effects of
quantum gravity. Besides, one of the important features of the
physics of the universe is its thermodynamic properties. In light of
this issue, we study the thermodynamic properties of the late time
universe by considering the gravitational effects (in the form of an
invariant infrared cutoff as a minimal measurable momentum),
starting from statistical physics. In this regard, we consider the
universe as a gaseous system and put it under the invariant infrared
cutoff effect. We show that the presence of this natural cutoff
leads to the accelerating expansion of the universe. It seems that
the infrared cutoff can be responsible for the late time
accelerating phase of the universe expansion.

As we know, the universe consists of different types of particles in
distinct degrees of freedom. In this paper, we study the late time
universe where the non-relativistic particles are dominant.
Therefore, under statistical physics, the distribution function in
the non-relativistic regime (for a thermodynamic system in
equilibrium) is characterized as follows
\begin{equation}\label{eq3}
f(\vec{p})=\frac{1}{e^\frac{E-\mu}{T}}\,.
\end{equation}
Note that, since in the late time universe we have $T\ll m$, we have
used approximation $e^{\frac{E-\mu}{T}}\pm1\approx
e^{\frac{E-\mu}{T}}$ in equation (\ref{eq3}) (we employ
Maxwell-Boltzmann statistics, i.e., we discard the term $\pm1$). We
also apply the non-relativistic prescription for particle energy
$E$. Henceforth, we call this energy ``visible energy". Our
motivation for this naming is to show that it is tangible as known
energy from known sources, which we will understand more about this
case in the following. Now, we obtain the number density $n$, energy
density $\rho$, and pressure ${\cal{P}}$ for various particles in
the late time universe with the above distribution function. This
means that particle distribution is performed as a function of phase
space. Therefore, to get the number density $n$, we integrate this
quantity over the momentum $p$
\begin{equation}\label{eq4}
n=\frac{g}{(2\pi)^3}\int f(\vec{p})d^3p\,,
\end{equation}
where $g$ express internal degrees of freedom and therefore the
density of states in the phase space is $\frac{g}{h^3}$. To obtain
the energy density $\rho$, we require to measure each state by its
energy $E=m+\frac{p^2}{2m}$ such that we get
\begin{equation}\label{eq5}
\rho=\frac{g}{(2\pi)^3}\int E(\vec{p})f(\vec{p})d^3p\,.
\end{equation}
Finally, the pressure $\cal{P}$ is specified as
\begin{equation}\label{eq6}
{\cal{P}}=\frac{g}{(2\pi)^3}\int
\frac{|\vec{p}|^2}{3E}f(\vec{p})d^3p\,,
\end{equation}

Now, we present some essential thermodynamic quantities which are
important in interpreting the accelerating expansion of the
universe. We will discuss them in subsequent sections. Accordingly,
we explore the thermodynamics of the late time universe by
incorporating quantum gravitational effects encoded in the Extended
Uncertainty Principle (EUP) that recognizes a minimal measurable
momentum. Since the late time universe is essentially an infrared
regime, the hypothesis of the equilibrium is applicable. We focus on
the thermodynamical quantities in the Maxwell-Boltzmann statistics
context in the presence of an IR cutoff due to EUP in the form of
$\Delta X\, \Delta P\geq \frac{\hbar}{2}\Big(1+\eta (\Delta
X)^{2}\Big)$.

At this point, we calculate $n$, $\rho$, and $p$ for
non-relativistic particles in the presence of infrared cutoff. We
find the number density as follows
\begin{eqnarray}\label{eq7}
n=\frac{4\pi g}{(2\pi)^3}\int_{P_{min}}^{\infty}p^2 e^\frac{-E}{T}dp
=\frac{g\textrm{e}^{-\frac{m}{T}}}{2\pi^2\xi^3}\bigg[\xi
P_{min}\textrm{e}^{-\frac{1}{2}\xi^2P_{min}^2}-\sqrt{\frac{\pi}{2}}\bigg(\textrm{erf}\Big(\frac{\xi
P_{min}}{\sqrt{2}}\Big)-1\bigg)\bigg]\,,
\end{eqnarray}
where $\xi=\frac{1+\eta x^2}{\sqrt{mT}}$ and $\textrm{erf}(x)$ is
the error function. Here it is necessary to point out again that, in
Eq. (\ref{eq2}), $x_i$ and $p_i$ are operators. In the definition of
$\xi$, $x$ is not an operator, but it is an eigenvalue of the
operator $x_i$. So, there is no inconsistency between the mentioned
relations. Note that in equation (\ref{eq7}), just for simplicity
and economy, we have introduced $\xi$ which contains the eigenvalue
$x$. The reason why we have not absorbed this additional term in
$P_{min}$ is that we want $P_{min}$ to be an invariant quantity
independent of $x$ and $T$. If we absorb this additional term in
$P_{min}$, we would be faced with a minimal momentum that depends on
temperature and position. This obviously breaks the invariance of
$P_{min}$.

In this work, we neglect the chemical potential ($\mu=0$), because
at the late time and in the non-relativistic limit where $T\ll m$,
this quantity is much smaller than the mass, that is, $\mu\ll m$
~\cite{Iglicki2018}.

The important point in these calculations is that the integration
range does not start from zero. In fact, due to the existence of the
minimal measurable momentum constraints in the spacetime structure,
the permissible range of these integrations is considered from
$P_{min}$ to infinity.

To determine the energy density, we apply $E\approx m$ and identify
from the definition in equation (\ref{eq5}) that
\begin{eqnarray}\label{eq8}
\rho=\frac{4\pi g}{(2\pi)^3}\int_{P_{min}}^{\infty}p^2E
e^\frac{-E}{T}dp
=\frac{mg\textrm{e}^{-\frac{m}{T}}}{2\pi^2\xi^3}\bigg[\xi
P_{min}\textrm{e}^{-\frac{1}{2}\xi^2P_{min}^2}-\sqrt{\frac{\pi}{2}}\bigg(\textrm{erf}\Big(\frac{\xi
P_{min}}{\sqrt{2}}\Big)-1\bigg)\bigg]\,.
\end{eqnarray}
This equation is also consistent with $\rho=mn$, where $n$ is given by equation (\ref{eq7}).\\

It should be noted that at the late time universe and in the
nonrelativistic limit, $p\ll m$ and therefore we can apply the
approximation $E\approx m$. Indeed, we apply this approximation just
for coefficient of the exponential term. Since in the exponential
term there is factor $\frac{E}{T}$, we are not allowed to apply the
approximation $p\ll m$, because the temperature is concerned with
the average kinetic energy as $T\approx\frac{p^2}{2m}$. Considering
that in the late time universe the temperature has a small value,
the expression $\frac{p^2}{2mT}$ is not small anymore. Therefore, we
cannot ignore this term in $e^\frac{-E}{T}$. That's why we are using
$E=m^2+\frac{p^2}{2m}$ in the exponential term. Note also that if we
discard the application of this approximation, the integrals cannot
be solved analytically and only numerical solutions are possible
(for example, the integral of Eq.(\ref{eq9})). In fact, without this
approximation, it is impossible to obtain an analytical solution for
the number density, the energy density, and the pressure as a
function of temperature. To find more details about this issue, see
Lecture $6$ in Ref.~\cite{Wrase}.

Finally, from equation (\ref{eq6}) we find the following expression
for the pressure
\begin{eqnarray}\label{eq9}
{\cal{P}}=\frac{4\pi g}{(2\pi)^3}\int_{P_{min}}^{\infty}p^4(1+\eta
x^2)^2 \frac{1}{3E} e^\frac{-E}{T}dp
=\frac{gT\textrm{e}^{-\frac{m}{T}}}{2\pi^2\xi^3}\bigg[\xi
P_{min}(1+\frac{1}{3}\xi^2P_{min}^2)\exp\Big(-\frac{1}{2}\xi^2P_{min}^2\Big)\hspace{1cm}
\nonumber\\
-\sqrt{\frac{\pi}{2}}\bigg(\textrm{erf}\Big(\frac{\xi
P_{min}}{\sqrt{2}}\Big)-1\bigg)\bigg]
=nT+\frac{gT}{6\pi^2}P_{min}^3\exp\Big(-\frac{1}{2}\xi^2P_{min}^2-\frac{m}{T}\Big)\,.
\end{eqnarray}
The re-scaled pressure of the non-relativistic particles versus
temperature, with a minimal momentum effect in the late time
universe, is plotted in figure 1. As the figure shows, in our model,
the effective pressure is negative which is an interesting and
favorite result. A remarkable issue about this result is that we got
a negative pressure without cosmological constant, scalar fields, or
modified gravity. In fact, it can be interpreted that the natural
infrared cutoff as a quantum gravitational effect itself has the
potential to describe the late time accelerating expansion of the
universe. We emphasize that this effect is achieved only with
quantum effects and standard model particles that have a well-known,
positive definite energy.

\begin{figure}
\begin{center}\includegraphics{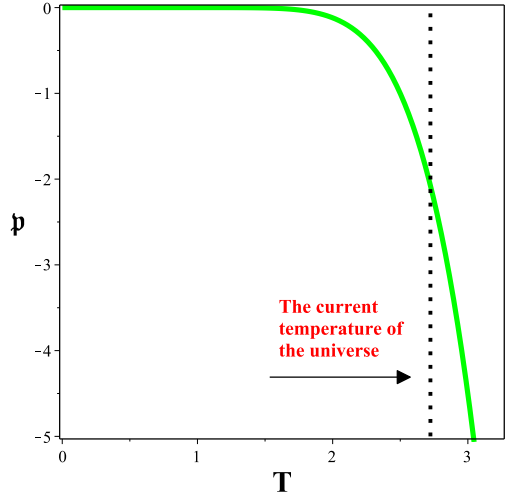}\vspace{5.5cm}
\end{center}
\caption{\label{fig1}\small {The re-scaled pressure of the
non-relativistic particles versus $T$ (in Kelvin) in the presence of
the minimal momentum at the late time universe, for
$\eta=-0.722\times10^{-47}$.}}
\end{figure}

\section{Cosmic Equation of State Parameter and Quantum Gravitational Effects}

The equation of state parameter for the standard, non-interacting,
non-relativistic particles (dust matter) is $w=0$. However, as we
have shown in previous sections, energy density and pressure are
modified in the presence of a minimal measurable momentum.
Therefore, the equation of state parameter of the standard
non-relativistic particle in the presence of quantum gravitational
effect takes the following form
\begin{eqnarray}\label{eq10}
w=\frac{p}{\rho} =\frac{\bigg(\Big[1+\frac{1}{3}\xi^2\Big]\xi
P_{min}\textrm{e}^{-\frac{\xi^2P_{min}^2}{2}}-\sqrt{\frac{\pi}{2}}\Big[\textrm{erf}(\frac{\xi
P_{min}}{\sqrt{2}})-1\Big]\bigg)T}{\bigg(\xi
P_{min}\textrm{e}^{-\frac{1}{2}\xi^2P_{min}^2}-\sqrt{\frac{\pi}{2}}\Big[\textrm{erf}(\frac{\xi
P_{min}}{\sqrt{2}})-1\Big]\bigg)m}\,.
\end{eqnarray}
As indicated, when the effects of quantum gravity are ignored
($\eta=0$), the equation of state parameter goes to $\frac{T}{m}$.
Given that in the late time universe $T\ll m$, so $w$ tends to zero
as in the standard model and as required. Moreover, the equation of
state parameter in the standard model is not exactly zero, but just
close to zero.

\begin{figure}
\begin{center}\includegraphics{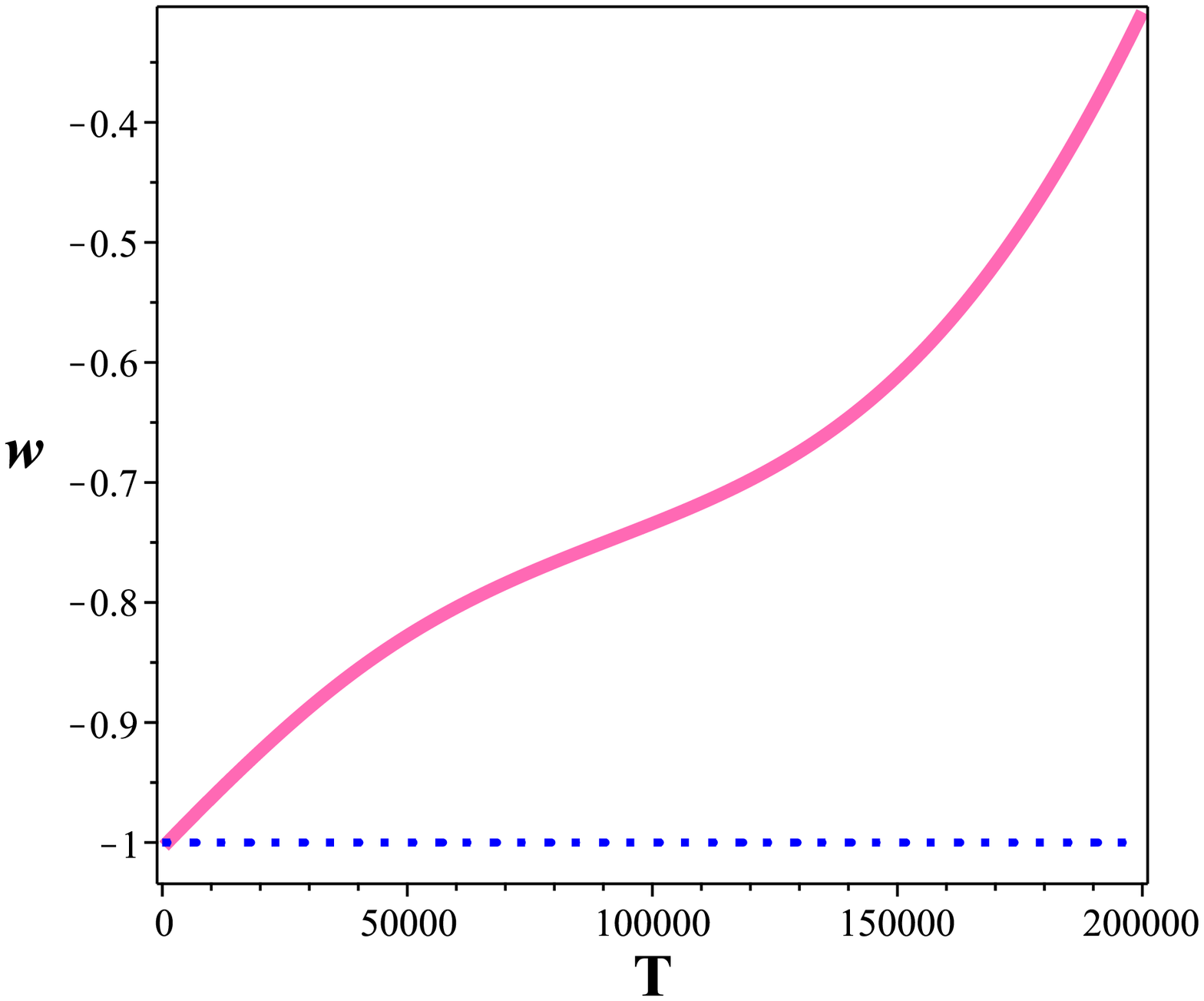} \includegraphics{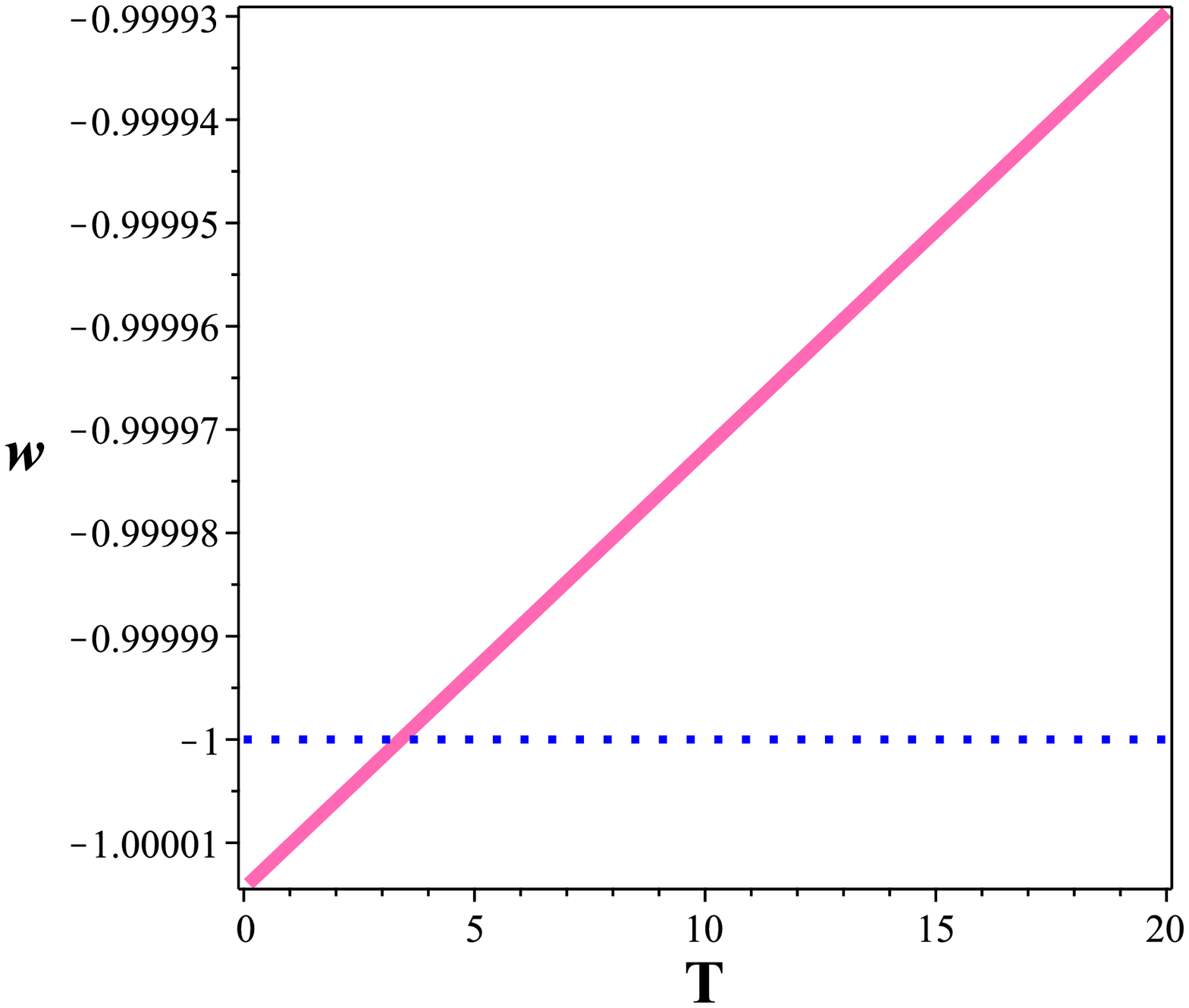} \vspace{5.5cm}
\end{center}
\caption{\label{fig2}\small {The equation of state parameter versus
temperature $T$ (in Kelvin) in the presence of the minimal momentum
for $\eta=-0.722\times10^{-47}$. In the right panel, we have zoomed
out the left panel to show the phantom line crossing more clearly.}}
\end{figure}

Now, we perform numerical analysis on the equation of state
parameter. In this regard, we adopt some sample values for the
$\eta$ parameter. Note that, both positive and negative values of
$\eta$ are physically viable. In our model, the negative values of
$\eta$ provide more interesting results. It is important to note
that in various scenarios of GUP in the form $[x,p]=i\hbar(1+\beta
p^2)$, the parameter $\beta$ can be negative~\cite{Scardigli2019}.
In this work, we have $[x,p]=i\hbar(1+\eta x^2)$, where
$\eta=\frac{\eta_0}{l_{pl}^2}$. By considering the negative values
of $\eta$, when $\eta_0x^2\rightarrow l_{pl}^2$, we get instantly
$[x,p]\rightarrow 0$. In other words, the negative values of $\eta$
can indicate the existence of a classical universe. Strictly
speaking, the classical trajectories in phase spaces of quantum
systems (including the universe) are the most probable in essence.
Accordingly, the negative values of $\eta$ allow us to have the
classical modes as well.

Figure 2 shows the behavior of the equation of state parameter
versus the temperature in the presence of the quantum gravitational
effects as a minimal measurable momentum. Crossing the phantom
divide line is also manifest in this plot for
$\eta=-0.722\times10^{-47}$. For this value of $\eta$, the phantom
line crossing occurs at $T=3.4\,K$. Also, the current value of $w$
in our setup is consistent with observational data
($w=-1.03\pm0.03$~\cite{Planck2018}). In fact, the behavior of the
equation of state parameter shows the accelerating expansion of the
universe in this model. Obviously, the effective role of the quantum
corrections of gravity has led to this interesting result, without
any vague justification. In other words, we were able to find an
alternative to dark energy, except that it is not dark but visible
alternative. In simple terms, visible energy described in this
framework is a substitute to dark energy without any darkness.

In figure 3, the phase space of $\eta$ and $T$ leading to
observationally viable values of the equation of state parameter is
shown. The shaded pink region in the plot shows the allowed values
of $\eta$ leading to $w=-1.03\pm0.03$~\cite{Planck2018}.

\begin{figure}
\begin{center}\includegraphics{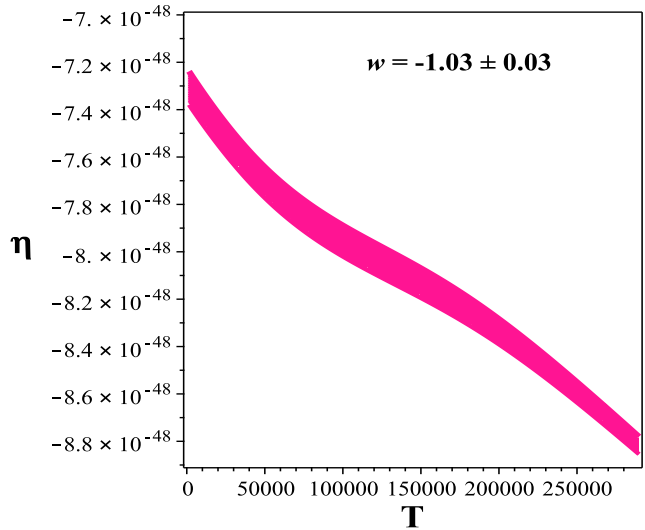}\vspace{5.5cm}
\end{center}
\caption{\label{fig3}\small {The phase space of $\eta$ and $T$ which
leads to the observationally viable values of the equation of state
parameter. The shaded pink region shows the allowed values of $\eta$
at the various times in cosmic history.}}
\end{figure}

\section{Conclusion}
Quantum gravity, though has not been well-formulated yet,
phenomenologically requires that the fabric of spacetime inherently
includes some natural cutoffs. As a result, the spacetime manifold
finds a discrete and lattice structure in quantum gravity regime.
Therefore, standard quantum mechanics under these natural cutoffs
should be automatically modified. These natural cutoffs are more
effectively depending on the scale of spacetime we are dealing with:
at short distances (Planck scales) one has a minimal measurable
length and at large distances (cosmological scales) one has a
minimal measurable momentum. In this paper, we have studied quantum
gravitational effects at large distances as a minimal measurable
momentum (natural infrared cutoff).

With this motivation, we have investigated thermodynamics of the
late time universe for non-relativistic particles in the presence of
a minimal measurable momentum. In this regard, we have considered a
gaseous system and applied certain quantum modifications to the
energy of these particles. We have called this energy as ``Visible
Energy" because this is tangible as known energy from known sources.
Then, consistent with statistical physics, the distribution function
in a non-relativistic framework has been corrected. Accordingly, we
have obtained some thermodynamic quantities, such as the number
density, energy density and pressure in this IR-modified setup. A
phenomenal issue about the modified pressure is that we have
obtained a negative pressure without cosmological constant, scalar
fields, or modified gravity. In fact, this result shows that the
natural infrared cutoff as a quantum gravitational effect can itself
describe the late time accelerating expansion of the universe. We
stress that this consequence is obtained only with quantum
gravitational effects and standard model particles that have
explicit and well-known energy. Finally, we have acquired a modified
equation of state parameter as an important cosmic quantity that
determines the behavior of the accelerating expansion of the
universe. It is so interesting that this important result has been
obtained just by taking quantum gravitation effects into account. In
fact, this framework provides a safe alternative for dark energy
without any darkness without recourse to some unknown components as
dark energy. In other words, our \emph{Visible Energy} is an
alternative to Dark Energy without \emph{darkness}. The strength of
our model is that we have described the accelerating expansion of
the universe and transition to the late time phantom phase with
\textit{standard particles} in the presence of infrared cutoff.
Finally we note that the value of the temperature and the equation
of state parameter in the time of transition to phantom phase are in
very good agreement with the known values from dark energy models
\cite{Nesseris2007}.

\end{document}